\documentclass[article,superscriptaddress]{revtex4}

\usepackage{float}
\usepackage{graphicx}

\newcommand{\Bk}{{\bf K}}

\newcommand{\bc}{{\bf c}}
\newcommand{\bd}{{\bf d}}

\newcommand{\br}{{\bf r}}

\newcommand{\bs}{{\bf s}}
\newcommand{\beq}{\begin{eqnarray}}
\newcommand{\eeq}{\end{eqnarray}}
\newcommand{\beqq}{\begin{eqnarray*}}
\newcommand{\eeqq}{\end{eqnarray*}}

\begin{document}

\title{Topological Crystalline Insulators in the SnTe Material Class}

\author{Timothy H. Hsieh}
\affiliation{Department of Physics, Massachusetts Institute of Technology, Cambridge, MA 02139}
\author{Hsin Lin}
\affiliation{Department of Physics, Northeastern University, Boston, Massachusetts 02115, USA}
\author{Junwei Liu}
\affiliation{Department of Physics, Massachusetts Institute of Technology, Cambridge, MA 02139}
\affiliation{Department of Physics and State Key Laboratory of Low-Dimensional Quantum
Physics,Tsinghua University, Beijing, 10084, China}
\author{Wenhui Duan}
\affiliation{Department of Physics and State Key Laboratory of Low-Dimensional Quantum
Physics,Tsinghua University, Beijing, 10084, China}
\author{Arun Bansil}
\affiliation{Department of Physics, Northeastern University, Boston, Massachusetts 02115, USA}
\author{Liang Fu}
\email{liangfu@mit.edu}
\affiliation{Department of Physics, Massachusetts Institute of Technology, Cambridge, MA 02139}

\maketitle


{\bf Topological crystalline insulators are new states of matter in which the topological nature of electronic structures arises from crystal 
symmetries.  Here we predict the first material realization of topological crystalline insulator in 
the semiconductor SnTe, by identifying its nonzero topological index.  
We predict that as a manifestation of this nontrivial topology, SnTe has metallic surface states with an even number of Dirac cones 
on high-symmetry crystal surfaces such as \{001\}, \{110\} and \{111\}. 
These surface states form a new type of high-mobility chiral electron gas, which is robust against disorder and topologically protected by reflection 
symmetry of the crystal with respect to \{110\} mirror plane. Breaking this mirror symmetry via elastic strain engineering 
or applying an in-plane magnetic field can open up a continuously tunable band gap on the surface, which may lead to wide-ranging applications 
in thermoelectrics, infrared detection, and tunable electronics.   
Closely related semiconductors PbTe and PbSe also become topological crystalline insulators after band inversion by pressure, strain and alloying.}

%
%

\section{Introduction}

The discovery of topological insulators\cite{kane, zhang, moore} has attracted much interest in topological states of matter beyond the existing 
$Z_2$ material class.  In searching for new phases, crystal symmetries play a multifaceted role by  either 
constraining the band topology\cite{fukane, bernevig, vishwanath} or 
engendering new ones\cite{magnetic, fu}. 
Topological crystalline insulators\cite{fu} are such new states of matter in which the topological nature of electronic structures arises from crystal 
symmetries. Thanks to the complexity and richness of crystal structures, the study of topological crystalline insulators has just began and    
a large number of topological crystalline insulators awaits discovery---both theoretically 
and experimentally.  
   
In this work, we theoretically demonstrate that SnTe is a topological crystalline insulator with mirror symmetry. 
As a consequence, it is predicted to have robust surface states with an {\it even} number of Dirac cones on crystal surfaces 
such as \{001\}, \{110\} or \{111\}, which are symmetric about \{110\} mirror planes. The notation \{hkl\} refers to the (hkl) plane and all those that are equivalent to them by virtue of the crystal symmetry. (A similar convention is used with directions:  $\langle n_1 n_2 n_3 \rangle$ refers to, collectively, the $[n_1n_2n_3]$ direction and its equivalent ones.) 
The existence of these surface states 
is dictated by a nonzero {\it integer} topological invariant---the mirror Chern number\cite{teofukane}.

\section{Results}

\subsection{Crystal Structure and Mirror Symmetry}

SnTe has a simple rocksalt structure (Fig.1a); its fundamental band gaps are located at four equivalent $L$ points  in the face-centered-cubic (FCC) Brillouin zone. 
It has long been established that the ordering of the conduction and valence bands  at $L$ points in SnTe 
is inverted {\it relative} to PbTe, so that the band gap of the alloy Pb$_{1-x}$Sn$_x$Te 
closes and re-opens as $x$ increases\cite{dms}.  Since this band inversion occurs at an even number of points, neither SnTe nor PbTe 
in the rocksalt structure is a topological insulator with $Z_2$ topological order. 

However, the above band inversion has significant consequences when the mirror symmetry of the FCC lattice is taken into consideration. 
Consider the plane $\Gamma L_1 L_2$ in momentum space, defined by the three points $\Gamma, L_1$ and $L_2$ (Fig.1b). 
Crystal momenta on this plane are invariant under reflection about the \{110\} mirror planes in real space.  This allows us to label the Bloch wavefunctions  on this plane by their eigenvalues $\pm i$ under the mirror operation $M$ (which satisfies $M^2=-1$ for spin $1/2$ electrons).
Each class of $M=\pm i$ mirror eigenstates has an associated Chern number $n_{\pm i}$, and the mirror Chern number $n_M$\cite{teofukane} is  defined by $n_M=(n_{+i} - n_{-i})/2$. Provided that mirror symmetry is present, $n_M$ is an integer topological invariant.  
A nonzero mirror Chern number defines a topological crystalline insulator with mirror symmetry.  

\subsection{Mirror Chern Numbers of SnTe and PbTe}

We first demonstrate that  SnTe and PbTe have mirror Chern numbers which differ by two, and hence, one of them is 
a topological crystalline insulator. 
This is established by considering the band inversion at four $L$ points between SnTe and PbTe. 
The $k \cdot p$ theory of the band structure near a given $L$ point is given by\cite{kp}:
\beq
H=  m \sigma_z + v (k_1 s_2 - k_2 s_1) \sigma_x + v_3 k_3 \sigma_y \label{H}.
\eeq  
 Here $k_1, k_2, k_3$ form an orthogonal system with $k_3$ along $\Gamma L$ and $k_1$ along the [110] direction perpendicular to the mirror plane; 
$\sigma_z=\pm 1$ corresponds to the $p$-orbital on the cation (Sn or Pb) and anion (Te) respectively; $s_3=\pm 1$ 
labels the total angular momentum $j=\pm 1/2$ along $\Gamma L$.  A positive $m$ means that the conduction and valence bands at $L$ are respectively derived from the cation and anion, and vice versa for negative $m$.  
The form of $H$ in Eq.(\ref{H}) is uniquely determined by the $D_{3d}$ point group symmetries which leave an $L$ point invariant\cite{fuberg}.
In particular,  reflection about the (110) mirror plane is represented by  
$
M = - i s_1
$. 
On the mirror-invariant plane $\Gamma L_1 L_2$ ($k_1=0$),  
$H$ reduces to 
$
H_0 = m \sigma_z -  v  k_2 s_1 \sigma_x + v_3 k_3 \sigma_y.  
$
Due to mirror symmetry, 
$H_0$ decomposes into the $s_1 = 1$ ($M=-i$) and $s_1=-1$ ($M=i$) subspaces: 
\beq
H_0^\pm = m \sigma_z \mp v k_2 \sigma_x + v_3 k_3 \sigma_y, 
\eeq
each of which describes a 2D massive Dirac fermion. 

In going from PbTe to SnTe (increasing $x$ in Pb$_{1-x}$Sn$_x$Te), the cation/anion character of the conduction/valence bands becomes switched at $L$, which in the $k\cdot p$ theory corresponds to $m\rightarrow -m$. 
The sign reversal of $m$ at one $L$ point changes the Chern number of the $s_1= \pm 1$ subspace by $\pm 1$, 
and hence the mirror Chern number changes by one. 
Furthermore, $L_1$ and $L_2$ are related by a two-fold rotation around the [1$\bar 1$0] axis. 
Under this rotation, both the spin polarization $s_1$ and the orientation of the $\Gamma L_1 L_2$ plane are flipped.  
As a result, the Berry curvatures at $L_1$ and $L_2$ are related by 
\beq
{\cal F}_{+i} (L_1)= - {\cal F}_{-i}(L_2), \; {\cal F}_{-i} (L_1) = -{\cal F}_{+i}(L_2). 
\eeq
so that 
\beq
({\cal F}_{+i} - {\cal F}_{-i}) |_{L_1} = ({\cal F}_{+i} - {\cal F}_{-i}) |_{L_2}, 
\eeq
i.e., both $L_1$ and $L_2$ contribute equally to the change in mirror Chern number.  
The net result is that the band inversion changes the mirror Chern number for the $\Gamma L_1 L_2$ plane by two.  
From this we deduce that SnTe and PbTe are topologically distinct phases, as long as the reflection symmetry with respect to any of 
the six equivalent \{110\} mirror planes is present.     

\subsection{SnTe is Topologically Nontrivial}

In order to determine which of the two is topologically nontrivial, 
we analyze their band structures obtained from first-principles density functional theory calculations\cite{wien2k, PAW,PAW2, VASP,VASP2,VASP3} 
using the generalized gradient approximation\cite{PBE}. 
The results we obtain are consistent with several previous studies\cite{PbTeGW0,PbTeAlloy, PbTePbSe, PbTe1985}. 
In particular, it is generally agreed that the band ordering at $L$ points is correct at the total-energy optimized volume. 

Based on two first-principles findings, 
we now demonstrate that, of the two materials, SnTe is the topologically nontrivial one.   
First, the conduction and valence bands of PbTe throughout the Brillouin zone are primarily 
derived from the $p$-orbitals of Pb and Te atoms respectively (Fig.2b). 
This suggests that PbTe is smoothly connected to the atomic limit, in which 
Pb orbitals are empty and Te orbitals filled due to their on-site energy difference. 
In contrast, the orbitals in the band structure of SnTe are switched near $L$ points:  
the conduction band edge is derived from Te and the valence band edge from Sn (Fig.2a). 
Therefore SnTe has an {\it intrinsically} inverted band structure ($m<0$). 
Our conclusion is further supported by the dependence of the band gap on the lattice constant obtained in 
first-principles calculations (Fig.2c) and  measured via pressure coefficients\cite{pressure}. 
As the lattice constant increases, the band gap of PbTe increases monotonically, whereas that of 
SnTe decreases to zero and then re-opens. This gap closing  signals a topological phase transition, in which 
SnTe at ambient pressure is a topological crystalline insulator with the aforementioned mirror Chern number $n_M=2$. 

\subsection{Surface States of SnTe}

The nonzero mirror Chern number in SnTe dictates the existence of surface states 
on any crystal surface which is symmetric about the \{110\} mirror planes. 
Three common surface terminations satisfying this condition are: 
\{001\} which is symmetric about \{110\} and \{1$\bar{1}$0\} mirror planes;
 \{111\} which is symmetric about \{1$\bar{1}$0\}, \{10$\bar{1}$\} and \{01$\bar{1}$\} mirror planes;
and \{110\} which is symmetric about \{1$\bar{1}$0\} mirror plane.    
According to the bulk-boundary correspondence, we now infer the topology of these surface bands.   
For the  \{001\}  surface,  the 
plane $\Gamma L_1 L_2$ in the bulk Brillouin zone projects onto the symmetry line  
$\bar{\Gamma} \bar{X_1}$ in the surface Brillouin zone, with both $L_1$ and $L_2$ projecting onto $\bar{X}_1$.  
The mirror Chern number $n_M=2$ dictates that  there must exist two {\it pairs} of 
counter-propagating, spin-polarized surface states with opposite mirror eigenvalues along the line $ \bar{X_1}- \bar{\Gamma} -\bar{X_1}$. 
By rotational symmetry, such surface states also appear  
along the line $\bar{X_2} - \bar{\Gamma} - \bar{X_2}$. But they are absent along any other mirror-invariant line. 
The crossing of  two mirror branches creates an anisotropic two-dimensional Dirac point with different velocities along the parallel and perpendicular direction.  
Therefore, the \{001\} surface states have 
four Dirac points located on the four equivalent $\bar{\Gamma} \bar{X}$ lines.  
Similar considerations apply to the other two surfaces.  For the $\{111\}$ surface, 
the  plane $\Gamma L_1 L_2$ projects onto the line  
$\bar{\Gamma} \bar{M}$. So there are two Dirac points along each of the three equivalent lines $\bar{\Gamma} \bar{M}$. 
For the $\{110\}$ surface, the  plane $\Gamma L_1 L_2$ projects onto the line $\bar{\Gamma} \bar{X}$, on which there are two Dirac points. 
In all cases, 
surface states of SnTe have an even number of  Dirac points, which can be easily distinguished from $Z_2$ topological insulators having an odd number. 
Our theoretical prediction of these  surface states  is the main result of this work. 

Using first-principles calculations, we now explicitly demonstrate the above surface states in a slab geometry 
along the [001] axis. Results for other surfaces will be published elsewhere. As predicted by the above topological band theory,   
two surface bands with opposite 
mirror eigenvalues are found to cross each other and form a Dirac point on the line $\bar \Gamma \bar X$ (Fig.3a). 
The Dirac velocity is found to be $1.7 \times 10^5$m/s in the $\bar \Gamma \bar X$ direction.  
Interestingly, these surface states exhibit a Lifshitz transition---a change of Fermi surface topology as a function of Fermi energy.
Fig.3b shows a set of Fermi surfaces at different energies.   
As the Fermi energy decreases from the Dirac point towards the valence band, the Fermi surface initially consists of 
two disconnected hole pockets outside $\bar{X}$; the two pockets then touch each other and reconnect to form a large hole and a  small electron pocket, both centered at $\bar{X}$.  

\section{DIscussion}

 In order to understand both the connection between the bulk and surface bands and the effect of potential perturbations at a microscopic level, 
we introduce a simplified tight-binding model for SnTe, detailed in Methods.  Using the tight-binding model, we now study the electronic properties of the \{001\}  surface states under various perturbations. Similar analysis applies to other surfaces.  
The doubly degenerate surface states $\psi_{\alpha=\pm}(\Bk_j)$ at the four Dirac points $\Bk_j$ have opposite mirror eigenvalues $i \alpha$, and 
hence opposite expectation values of spin polarization {\it perpendicular} to the ${\bar \Gamma} \Bk_j$ direction.       
 For convenience, we choose a natural basis in which the relative phases between the wavefunctions at different Dirac points are fixed by the four-fold rotation relating them
\beq
C_4: &\;&  \psi_\alpha (\Bk_j) \rightarrow e^{-i \alpha \pi/4} \psi_\alpha(\Bk_{j+1}). \label{sym}
\eeq
In this basis,  the  $k \cdot p$ Hamiltonians at  four Dirac points take an  identical form: 
\beq
H_{\rm sf} =  v_\perp k_1 s_2 - v_{\parallel} k_2 s_1, \label{surface}
\eeq 
Here  $k_1$ and $k_2$ form a {\it local} right-handed coordinate system centered at each $\Bk_j$, with $k_2$ parallel to $\Bk_j$. 

It is important to note that the four branches of surface Dirac fermions have the {\it same} chirality, defined by the relative sign of $v_\perp$ and $v_\parallel$ in (\ref{surface}).  
As such, the surface states of the topological crystalline insulator SnTe 
form a ``chiral'' Dirac metal protected by crystal symmetries, thereby defining a new symmetry/topology universality class.  
In particular, provided that mirror symmetry is present ($k_1 \rightarrow -k_1, s_2 \rightarrow -s_2$), the Dirac points here can move along the $\bar \Gamma \bar X$ line but cannot 
annihilate with each other. 
 
We now consider ways to engineer a band gap on the surface. 
Perturbations which break mirror symmetries can generate Dirac mass term $m_j s_3$ and 
thus open up energy gaps $E_j=2|m_j|$ at the Dirac points. 
The nature of the gapped phase depends on the relative signs of $m_j$, which is determined by 
the symmetry of the perturbation. 
For example, a perpendicular magnetic field $B$ couples to electron's spin 
and yields Dirac masses of the same sign due to rotational invariance: $m_1=m_2=m_3=m_4\propto B$. This drives the surface states into an integer quantum Hall state. 

Those perturbations which break the four-fold rotation symmetry of the \{001\} surface are  particularly interesting. 
One example is a structure distortion with atoms being displaced by a vector $\bf{u}$, which can be introduced by strain and possibly external electric field. 
Also, SnTe is known to spontaneously distort along the $\langle 111 \rangle$ direction into a rhombohedral structure at low temperature.  
This perturbation is time-reversal invariant but breaks rotational symmetry, leading to a ferroelectric phase.   
While this distortion has negligible effect on the band structure in the bulk, it can dramatically affect the Dirac surface states.  
The effect of the distortion can be captured by adding a modulated hopping term to the tight binding model (see Methods).
By symmetry analysis (\ref{sym}), one can show that the distortion gives the following anisotropic Dirac mass terms: 
\beq
m_j \propto ( {\bf u} \times {\Bk}_j ) \cdot {\hat z}, \label{mass}
\eeq
where $\hat{z}$ is the surface normal. As a result, the metallic surface acquires a band gap, which is {\it linearly} 
proportional to the magnitude of structural distortion and depends on its direction. 
For the rhombohedral distortion ${\bf u} \propto (111)$,  the two Dirac points along the $(1\bar{1}0)$ direction are gapped, 
but the other two along the (110) direction remain gapless due to the unbroken (110) mirror plane. 
The ability to continuously tune the surface band gap via applying strain 
suggests potential electronic and optoelectronic device applications based on topological crystalline insulators such as SnTe.  

Likewise, an in-plane magnetic field in the $\bf u$ direction generates Dirac mass terms given by $m_j \propto  {\bf u} \cdot {\Bk}_j$. 
In particular, when the magnetic field is perpendicular to one of the mirror planes, the system is still symmetric about this 
mirror plane and hence continues to exhibit gapless surface states despite the broken time reversal symmetry.  
Therefore topological crystalline insulators defined by mirror Chern number also exist in magnetic systems such as Mn- and Cr-doped SnTe\cite{story}, 
contrary to existing $Z_2$ topological insulators.   

As Eq.(\ref{mass}) shows,  when the displacement vector $\bf u$  
points along the mirror invariant line ${\bar \Gamma} {\bar X}$, the surface states remain gapless at two Dirac points. 
Therefore, there are four distinct types of fully-gapped surface ferroelectric phases depending on the direction of ${\bf u}$ (Fig.4). These ferroelectric phases 
correspond to breaking crystal symmetry  in ``different directions'' and their electronic structures cannot be adiabatically connected. 
Now consider two such gapped phases  (A and B), which are spatially adjacent to each other. 
The domain wall between them turns out to be particularly interesting if ${\bf u}_A$ lies in the I-III quadrants and ${\bf u}_B$ the II-IV quadrants. 
Due to their different displacement vectors, a pair of Dirac masses at opposite momenta change sign in going from A to B.   
As a result, there exists a time-reversed pair of counter-propagating 1D gapless states bound to the domain wall. 
Similar to the edge states of quantum spin Hall insulators, 
the domain wall states here are protected from elastic backscattering\cite{km}, and form perfectly conducting channels. 

The existence of such domain wall states explains the robustness of these surface states against time-reversal-invariant disorder. By definition, disorder
breaks mirror symmetry in a random way locally, not macroscopically.    
For the sake of argument, let us assume that the disorder potential is slowly varying. The surface is then an equal-weight mixture of the aforementioned 
type I-III  and type II-IV ferroelectric domains, with domain walls percolating throughout the surface. 
Since the domain wall is a perfectly conducting channel,  the entire surface must be conducting.  
A similar situation occurs in disordered weak topological insulators\cite{stern, halperin, liu, mong}. 

The above analysis of disorder also accounts for surface roughness. Surfaces without any symmetry are generically gapped. 
Similar to structural distortions, different crystal faces correspond to breaking crystal symmetry in different directions. By the same reasoning, 
the aforementioned 1D domain wall states exist on edges connecting them. As a result, a surface 
is guaranteed to be metallic, if it is 
mirror symmetric {\it on average} in the sense of having the same Bragg diffraction peak as the clean surface.    
A comprehensive theory of disordered topological crystalline insulators is in progress.


On the experimental side, 
the surface states of SnTe we predicted can be readily detected in angle-resolved photoemission spectroscopy (ARPES) and 
tunneling spectroscopy experiments. 
The underlying mirror Chern number can be deduced from the spin polarization of surface states\cite{hsieh}. 
Moreover, SnTe-based thin films and superlattices have remarkably high mobility exceeding $2500$cm$^2$/Vs at room temperature\cite{sntemobility1, sntemobility2}, 
which provide a promising platform for device applications.  

Finally, we relate our work to a wider class of materials, including PbTe and PbSe. 
While both are topologically trivial at ambient pressure, 
our first-principles calculation (Fig.2c) shows that decreasing the lattice constant by 2\% 
inverts the band gap and drives PbTe into a topological crystalline insulator.   
This band inversion is realized under moderate pressure (around 3GPa in PbTe and 2GPa in PbSe
\cite{pressure,PbTePbSe}). Alternatively, one can achieve the topological regime  by 
growing these materials on substrates with smaller lattice constants. As a precedent, 
high-quality PbTe quantum wells have been fabricated and exhibit ballistic transport\cite{conductance, qhe}. 
Moreover,  the alloys Pb$_{1-x}$Sn$_x$Te and Pb$_{1-x}$Sn$_x$Se undergo 
band inversion as Sn composition increases\cite{pbsnse1, pbsnse2}, and hence become topological crystalline insulators on the inverted side.

We briefly comment on how our work relates to early pioneering field-theoretic studies,  which predicted the existence of  
two dimensional massless Dirac fermions at the interface  of  PbTe and SnTe\cite{volkov, drew}, 
or domain wall of PbTe\cite{fradkin}. Our work has made it clear that only interfaces symmetric about the \{110\} mirror plane have 
protected gapless states, which are solely derived from the topological crystalline insulator SnTe and exist  even when PbTe is removed. 
In light of ther topological nature which we identified,  SnTe material class in IV-VI semiconductors is likely to 
lead a new generation of topological materials. 
  
\section{Methods}

The tight binding model for SnTe is constructed from the Wannier functions of the conduction and valence bands, 
which are primarily three $p$-orbitals of Sn and Te atoms. The Hamiltonian $H_{tb}$ is given by  
\beq
H_{tb} &=&  m  \sum_{j}(-1)^j   \sum_{\br, \alpha}     \bc^\dagger_{j\alpha}(\br) \cdot \bc_{j\alpha}(\br)   \nonumber \\
&+& \sum_{j, j'} t_{jj'}  \sum_{(\br, \br'), \alpha }  \bc^\dagger_{j\alpha}(\br) \cdot {\hat d}_{\br\br'}  \; {\hat d}_{\br\br'}  \cdot \bc_{j'\alpha}(\br')  + h.c. \nonumber \\
&+& \sum_j   i \lambda_j \sum_{\br, \alpha,\beta}    \bc^\dagger_{j\alpha}(\br) \times \bc_{j\beta}(\br)  \cdot \bs_{\alpha \beta}.  \label{tb}
\eeq 
Here $\br$ labels the site, $j=1,2$ labels the Sn or Te atom, $\alpha=\uparrow, \downarrow$ labels electron's spin.    
The components of vectors $\bc^\dagger$ and $\bc$ correspond to the three $p$-orbitals. 
In the Hamiltonian (\ref{tb}), $m$ is  the on-site potential difference between Sn and Te;  
$ t_{12}=t_{21}$ is the nearest-neighbor hopping amplitude between Sn and Te; 
$t_{11}$ and $t_{22}$ are the  next nearest-neighbor hopping amplitudes within a sublattice; ${\hat d}_{\br\br'}$ is the unit vector connecting site  $\br$ to $\br'$.  
The second line thus represents $\sigma$-bond hopping (head to tail) of the $p$-orbitals.
The $\lambda_{1, 2}$ term is  ${\bf L} \cdot {\bf s}$ atomic spin-orbit coupling, where $L_j = i \epsilon_{jkl}$ is the orbital angular momentum in $p$-orbital basis. 
The bulk and surface bands of the above tight-binding Hamiltonian nicely reproduce the essential features of the first-principles calculation, and additional terms 
such as $\pi$-bond hopping of the $p$-orbitals can be added to improve the fit.    

The effect of a structural distortion in which atoms are displaced by {\bf u} can be captured by adding the modulated hopping
\beq
\delta t  \sum_{(\br, \br'), \alpha }  \bc^\dagger_{1\alpha}(\br) \cdot {\bd}_{\br\br'}  \; {\bf u} \cdot \bc_{2\alpha}(\br') + h.c.
\eeq 
to the tight binding model.

\section{Author Contributions}
T.H. and L.F. made theoretical analysis of SnTe and related IV-VI semiconductors. H.L., J.L., W.D. and A.B. performed ab-initio calculations. 
L.F. conceived the idea for topological crystalline insulators in the SnTe material class, supervised the whole project, and wrote part of the manuscript with 
contributions from all authors. 

Correspondence should be addressed to L. F. (liangfu@mit.edu)

\section{Acknowledgement}
 
This work is supported by NSF Graduate Research Fellowship No. 0645960 (TH) and
start-up funds from MIT (LF). The work at Northeastern University is supported by the Division of
Materials Science and Engineering, Basic Energy Sciences, US
Department of Energy through grant number DE-FG02-07ER46352
and benefitted from the allocation of supercomputer time at
NERSC and Northeastern University's Advanced Scientific Computation Center.
JL and WD acknowlege the financial support from Ministry of Science and Technology of China (Nos. 2011CB921901 and
2011CB606405), the National Natural Science Foundation of China and NSF Grant No. DMR-1005541.

 \begin{figure}
\centering
\includegraphics[height=1.8in]{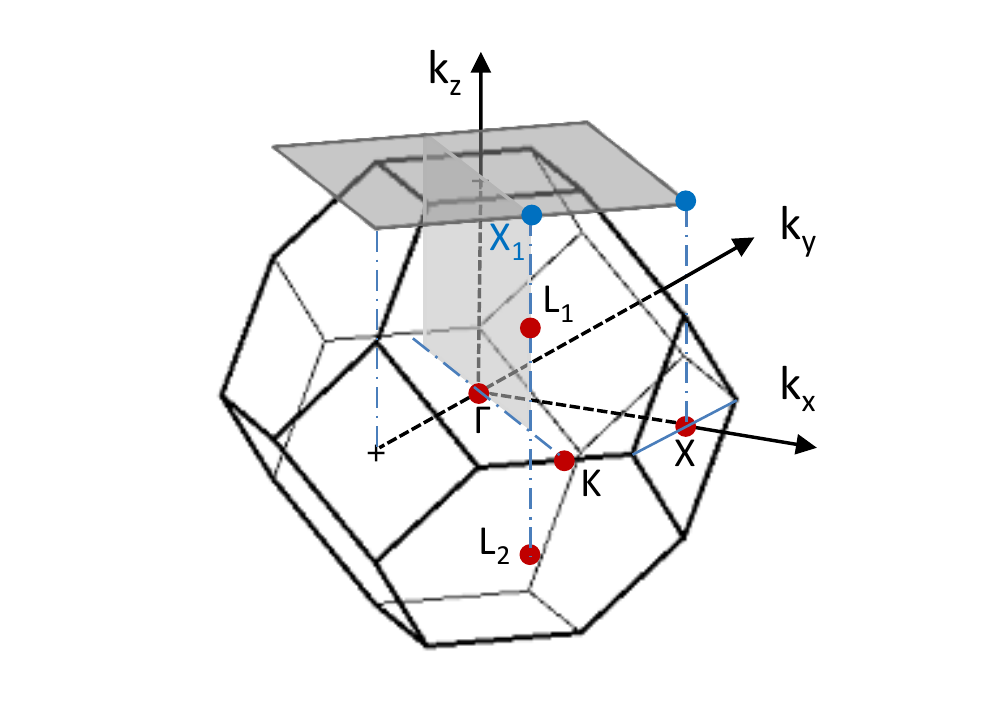}
\caption{ {\bf SnTe lattice and Brillouin zone}: (a) the crystal structure of SnTe; (b) the FCC Brillouin zone showing the 
plane $\Gamma L_1 L_2$ which is invariant under reflection about the (110) axis and projects onto 
the $\bar{\Gamma} \bar{X_1}$ line in the [001] surface. }
\end{figure}

\newpage

\begin{figure}
\centering
\includegraphics[width=3in, height=3.5in]{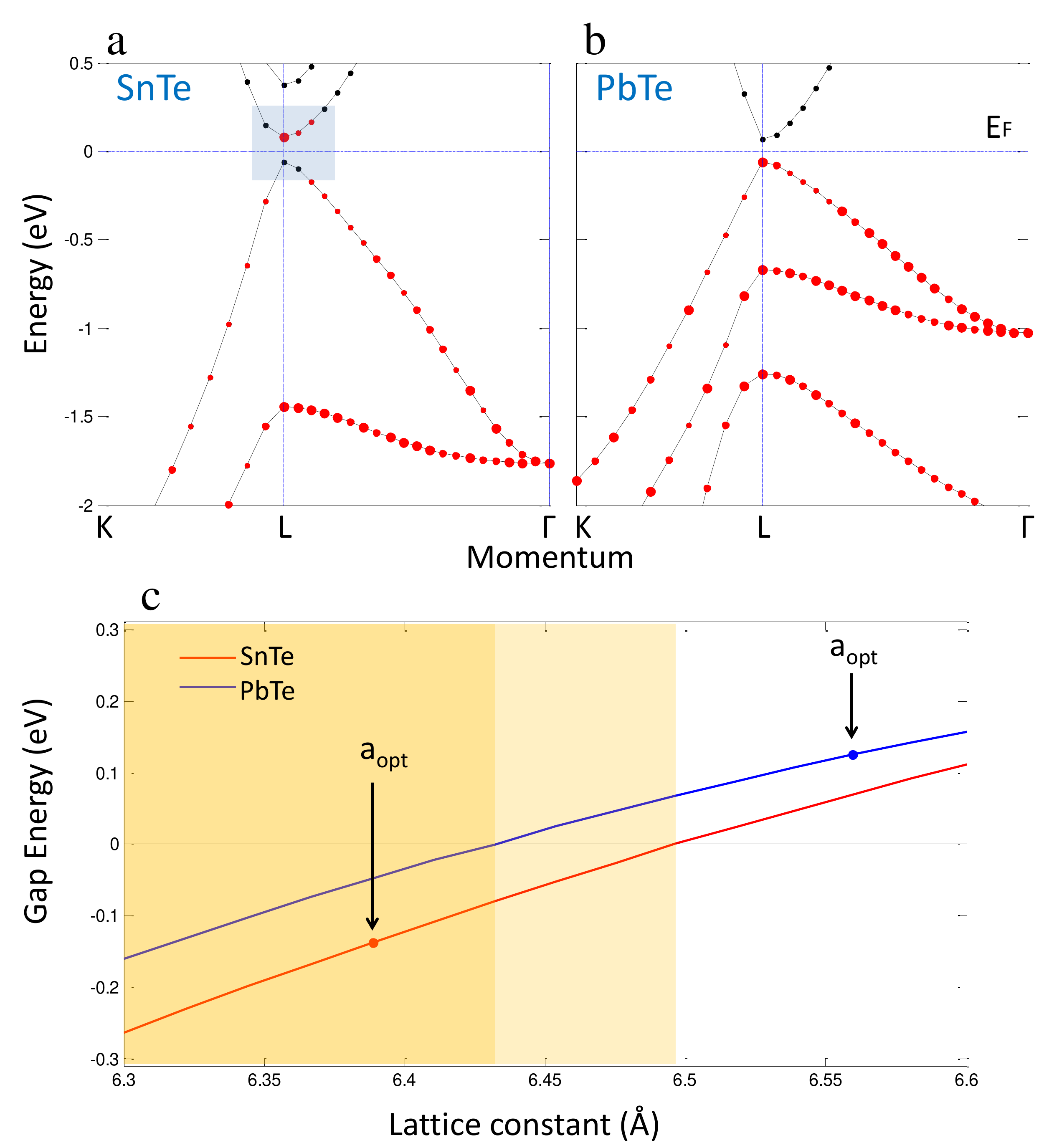}
\caption{{\bf Electronic structures and band-gap evolution of SnTe and PbTe}:  
The electronic band structures of SnTe and PbTe are shown in a and b, respectively.
The size of the red dots represents the fraction of electronic charge residing on Te atoms. 
The exchange of the band character at L point as highlighted in the grey area indicates the {\it intrinsic} band inversion of SnTe.
The band gap energy as a function of the lattice constants is shown in c. 
The negative gap area indicates the topological crystalline insulator phase. }
\end{figure}

\begin{figure}
\centering
\includegraphics[width=3in]{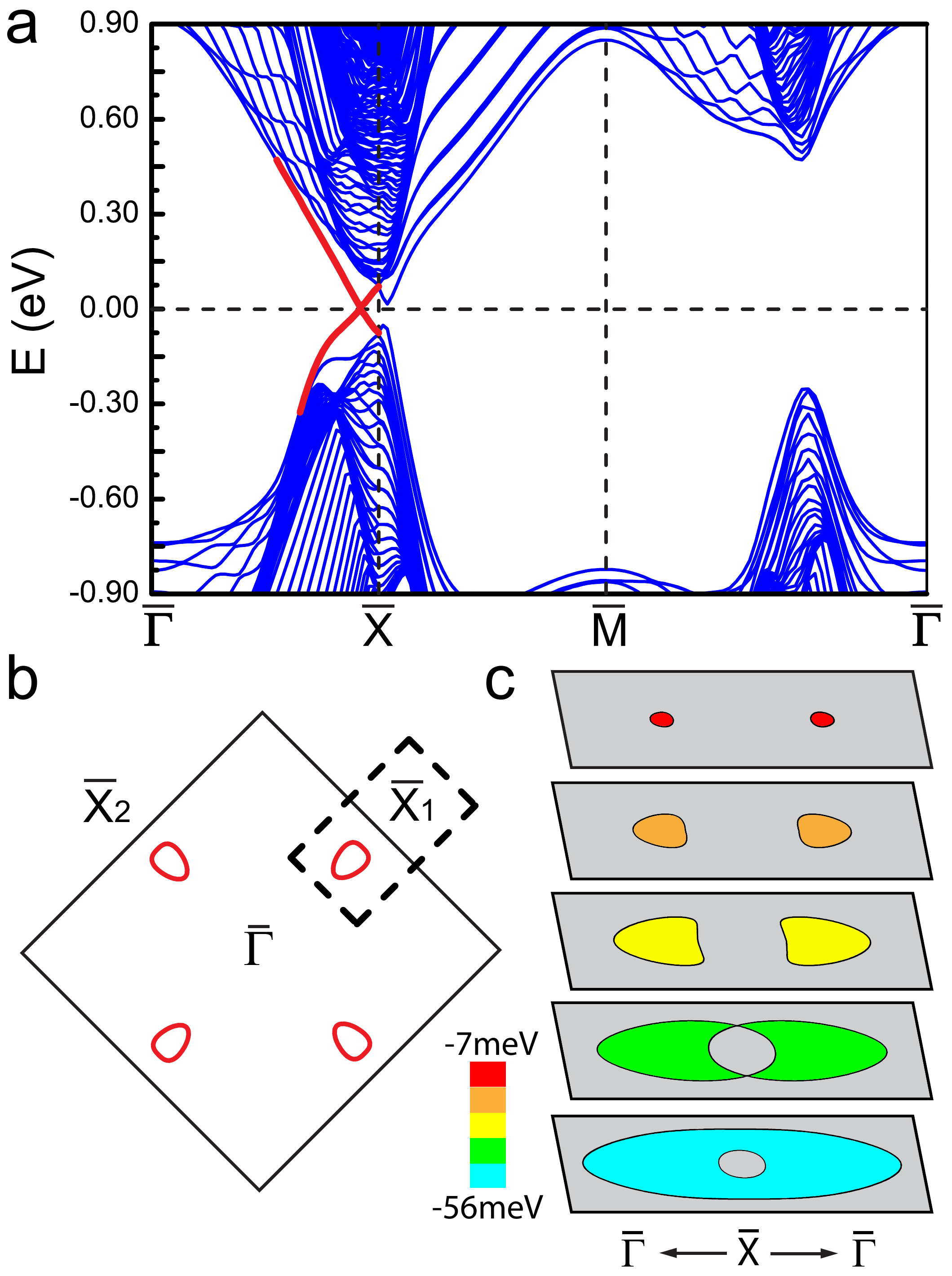}
\caption{{\bf The [001] surface states of SnTe}: (a) band dispersion and (b) Fermi surface. Note that 
  in between $\bar{\Gamma}$ and $\bar{X}$ two surface bands with opposite mirror eigenvalues cross the Fermi energy, in 
  agreement with $n_M=2$. (c) a set of Fermi surfaces at different energies, exhibiting a Lifshitz transition.}
\end{figure}

\begin{figure}
\centering
\includegraphics[width=3in]{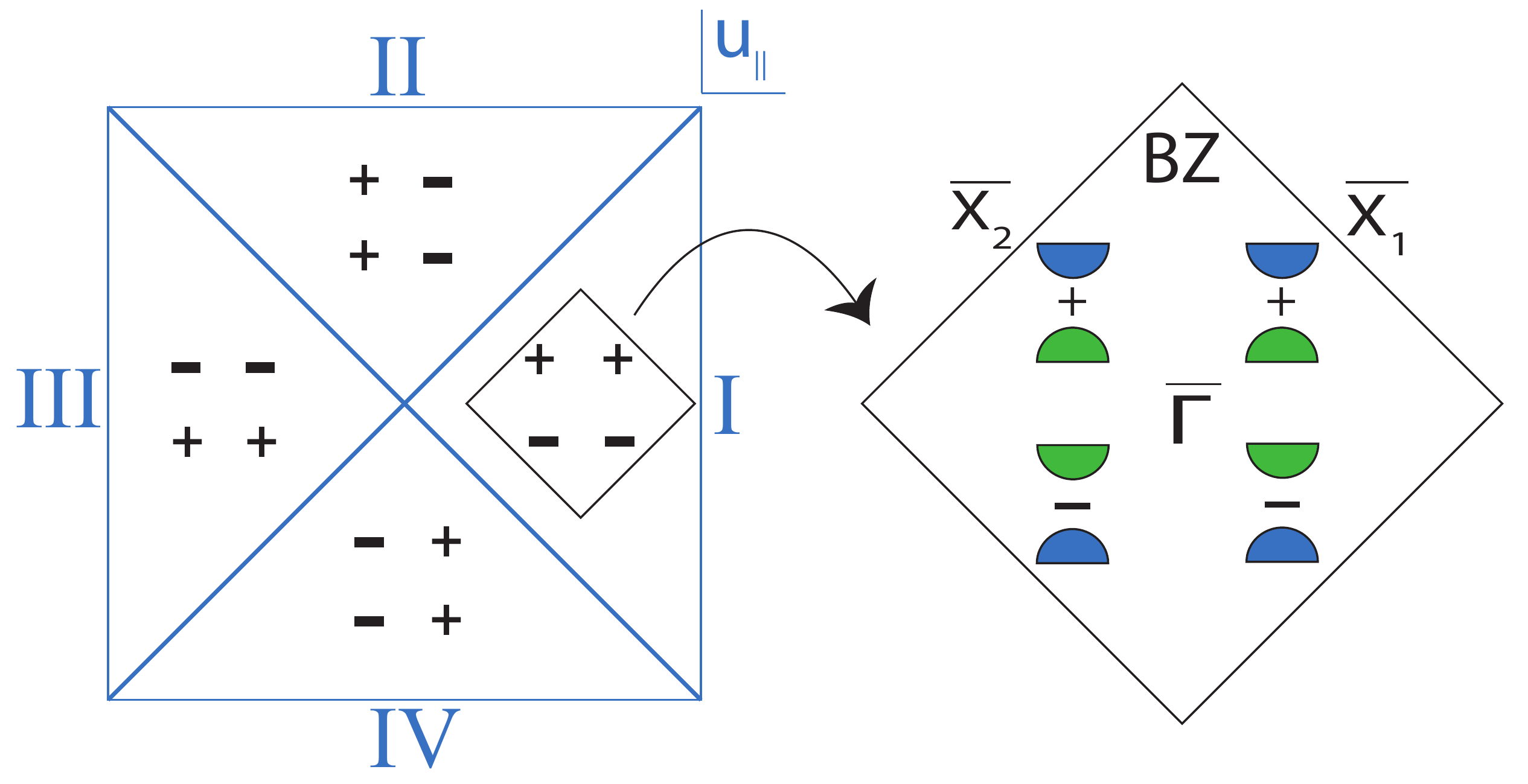}
\caption{{\bf The four surface ferroelectric phases, each with the signs of the Dirac masses}: 
Under a ferroelectric distortion with a displacement ${\bf u}$, the four surface Dirac points (detailed on the right) 
are gapped with mass signs depending on the direction of ${\bf u}_\parallel$.  This leads to the four gapped phases (I-IV).}
\end{figure}


\begin{thebibliography}{10}

\bibitem{kane} 
Hasan, M. and Kane, C.L. Colloquium: Topological insulators. \emph{Rev. Mod. Phys} {\bf 82}, 3045 (2010). 

\bibitem{zhang} 
Qi, X.L. and Zhang, S.C. Topological insulators and superconductors. \emph{Rev. Mod. Phys.} {\bf 83}, 1057 (2011). 

\bibitem{moore} 
Moore, J.E. The birth of topological insulators. \emph{Nature} {\bf 464}, 194 (2010).

\bibitem{fukane} 
Fu, L. and Kane, C.L. Topological insulators with inversion symmetry. \emph{Phys. Rev. B} {\bf 76}, 045302 (2007).

\bibitem{vishwanath} 
Turner, A. M. , Zhang, Y. and Vishwanath, A. Entanglement and inversion symmetry in topological insulators. \emph{Phys. Rev. B} {\bf 82}, 241102(R) (2010). 

\bibitem{bernevig} 
Hughes, T. L. , Prodan, E., and Bernevig, B.A. Inversion-symmetric topological insulators. \emph{Phys. Rev. B} {\bf 83}, 245132 (2011).

\bibitem{magnetic}
Mong, R. K. , Essin, A. M., and Moore, J.E. Antiferromagnetic topological insulators. \emph{Phys. Rev. B} \textbf{81}, 245209 (2010).


\bibitem{fu}
Fu, L. Topological crystalline insulators. \emph{Phys. Rev. Lett.} {\bf 106}, 106802 (2011).


\bibitem{teofukane}
Teo, J. C. Y. , Fu, L., and Kane, C.L. Surface states and topological invariants in three-dimensional topological insulators: Application to $Bi_{1?x}Sb_x$. \emph{Phys. Rev. B} {\bf 78}, 045426 (2007).

\bibitem{dms}
Dimmock, J. O. , Melngailis, I., and Strauss, A. J. Band Structure and Laser Action in $Pb_xSn_{1-x}Te$. \emph{Phys. Rev. Lett.} {\bf 26}, 1193 (1966).

\bibitem{kp}
Mitchell, D. L. and Wallis, R. F. Theoretical Energy-Band Parameters for the Lead Salts. \emph{Phys. Rev.} 151, 581Ð595 (1966). 

\bibitem{fuberg}
Fu, L. and Berg, E. Odd-Parity Topological Superconductors: Theory and Application to $Cu_xBi_2Se_3$. \emph{Phys. Rev. Lett.}{\bf 105}, 097001 (2010).

\bibitem{wien2k}Blaha, P.
Schwarz, K., Madsen, G. K. H., Kvasnicka, D. $\&$ Luitz, J.
\textit{WIEN2k, An Augmented Plane Wave Plus Local Orbitals Program for
Calculating Crystal Properties.}
(Karlheinz Schwarz, Techn. University Wien, Austria, 2001).
%
\bibitem{PAW}
Bl$\ddot{o}$chl, P. E.  Projector augmented-wave method. \emph{Phys. Rev. B.} {\bf 50}, 17953 (1994).
%
\bibitem{PAW2}
Kresse, G. and Joubert, J.  From ultrasoft pseudopotentials to the projector augmented-wave method. \emph{Phys. Rev. B.} {\bf 59}, 1758 (1999).
%
\bibitem{VASP}
Kresse, G. and Hafner, J. Ab initio molecular dynamics for open-shell transition metals. \emph{Phys. Rev. B.} {\bf 48}, 13115 (1993).
%
\bibitem{VASP2}
Kresse, G. and Furthm$\ddot{u}$ller, J. Efficiency of ab-initio total energy calculations for metals and semiconductors using a plane-wave basis set. \emph{Comput. Mater. Sci.} {\bf 6}, 15 (1996).
%
\bibitem{VASP3}
Kresse, G. and Furthm$\ddot{u}$ller, J.  Efficient iterative schemes for ab initio total-energy calculations using a plane-wave basis set. \emph{Phys. Rev. B.} {\bf 54}, 11169 (1996).

\bibitem{PBE} Perdew, J. P., Burke, K. \& Ernzerhof, M. Generalized Gradient Approximation made simple,  {\it Phys. Rev. Lett.} \textbf{77} 3865-3868 (1996).
%



\bibitem{PbTeGW0}
Hummer, K., Grüneis, A., and Kresse, G. Structural and electronic properties of lead chalcogenides from first principles. \emph{Phys. Rev. B} {\bf 75}, 195211  (2007).

\bibitem{PbTeAlloy}
X. Gao and M. S. Daw, Phys. Rev. B {\bf 77}, 033103  (2008).

\bibitem{PbTePbSe}
Svane, A. et al.
Quasiparticle self-consistent GW calculations for PbS, PbSe, and PbTe: Band structure and pressure coefficients.
\emph{Phys. Rev. B} {\bf 81}, 245120   (2010).


\bibitem{PbTe1985}
Rabe, K. M. \& Joannopoulos, J. D.
Ab initio relativistic pseudopotential study of the zero-temperature structural properties of SnTe and PbTe
\emph{Phys. Rev. B} {\bf 32}, 2302  (1985).

\bibitem{pressure}
Shchennikov, V. V. \& Ovsyannikov, S. V.
Thermoelectric power, magnetoresistance of lead chalcogenides in the region of phase transitions under pressure.
\emph{Solid State Commun.} {\bf 126}, 373 (2003).

 
%
%

\bibitem{story}
T. Story, R. R. Galazka, R. B. Frankel \& P. A. Wolff, Carrier-concentrationÐinduced ferromagnetism in PbSnMnTe
\emph{Phys. Rev. Lett.} {\bf 56}, 777 (1986).


\bibitem{km}
Kane, C. L. \& Mele, E. J.
Quantum Spin Hall Effect in Graphene.
\emph{Phys. Rev. Lett.} {\bf 95}, 226801 (2005).

 


\bibitem{stern}
Ringel, Z., Kraus, Y. E., \& Stern, A.
The strong side of weak topological insulators.
Preprint at arXiv:1105.4351 (2011). 

\bibitem{halperin}
Halperin, B. I., private communication

\bibitem{liu}
Liu, C. X., Qi, X. L. \& Zhang, S. C.
Half quantum spin Hall effect on the surface of weak topological insulators.
\emph{Physica E}, {\bf 44}, 906 (2012).


\bibitem{mong}
Mong, R. S. K., Bardarson, J. H., \& Moore, J. E.
Quantum transport and two-parameter scaling at the surface of a weak topological insulator.
\emph{Phys. Rev. Lett.} {\bf 108}, 076804 (2012).





%

\bibitem{hsieh}
Hsieh, D. et al,
Observation of Unconventional Quantum Spin Textures in Topological Insulators.
\emph{Science}, {\bf 323}, 5916 (2009).

\bibitem{sntemobility1}
Ishida, A. et al.
Electrical and thermoelectrical properties of SnTe-based films and superlattices.
\emph{Appl. Phys. Lett.} {\bf 95}, 122106 (2009).

\bibitem{sntemobility2}
Ishida, A. et al.
Electrical and optical properties of SnEuTe and SnSrTe films.
\emph{J. Appl. Phys.} {\bf 107}, 123708 (2010).


\bibitem{conductance}
Grabecki, G. et al.
PbTe - A new medium for quantum ballistic devices.
\emph{Physica E}, {\bf 34}, 560 (2006).

\bibitem{qhe}
Chitta, V. A. et al
Multivalley transport and the integer quantum Hall effect in a PbTe quantum well.
\emph{Phys. Rev. B} {\bf 72}, 195326 (2005).

\bibitem{pbsnse1}
Strauss, A. J. 
Inversion of Conduction and Valence Bands in Pb$_{1-x}$Sn$_x$Se Alloys. 
\emph{Phys. Rev} {\bf 157}, 608 (1967)

\bibitem{pbsnse2}
Calawa, A. R. et al. 
Magnetic Field Dependence of Laser Emission in Pb$_{1-x}$Sn$_x$Se Diodes
\emph{Phys. Rev. Lett} {\bf 23}, 7 (1969).

\bibitem{volkov}
Volkov, B.A. \& Pankratov, O.A. 
\emph{Pisma Zh. Eksp. Teor. Fiz.} 42, 145-148 (1985); English transl. JETP Lett. 42, 178-181 (1985).

\bibitem{drew}
Korenman, K. \& Drew, H. D.
Subbands in the gap in inverted-band semiconductor quantum wells.
\emph{Phys. Rev. B} {\bf 35}, 6446 (1987).

\bibitem{fradkin}
Fradkin, E., Dagotto, E., \& Boyanovsky, D.
Physical Realization of the Parity Anomaly in Condensed Matter Physics.
\emph{Phys. Rev. Lett.} {\bf 57}, 2967 (1986).


\end{thebibliography}
\end{document}